\documentclass[onecolumn]{aastex62}
\title{Constraining Spatial Densities of Early Ice Formation in Small Dense Molecular Cores from Extinction Mapping}

\newcommand\aastex{AAS\TeX}

\newcommand{\choh}{CH$_3$OH }
\newcommand{\hto}{H$_2$O }

\newcommand{\Av}{A$_V$ }
\newcommand{\Ak}{A$_K$ }
\newcommand{\cm}{cm$^{-3}$ }
\usepackage{xcolor}
\usepackage{url}
\usepackage{longtable} 
\usepackage{times}
\usepackage{float}
\usepackage{wrapfig}
\usepackage{rotating}
\usepackage{amssymb}
\usepackage{xcolor}
\usepackage{dejavu}
\usepackage{colortbl,booktabs,array,tabularx}
\usepackage{lipsum,graphicx,float}
\usepackage[caption=false]{subfig}
\usepackage{graphicx}
\usepackage{rotating}
\usepackage{scrextend}
\usepackage{hyperref}

\newcommand{\Msun}{\ensuremath{M_\odot}}

\accepted{June 11, 2021}

\submitjournal{Astrophysical Journal}

\shorttitle{\aastex\ Extinction Maps}
\shortauthors{Chu et al.}

\begin{document}

\title{Constraining Spatial Densities of Early Ice Formation in Small Dense Molecular Cores from Extinction Maps}

\correspondingauthor{Laurie E U Chu}
\email{lauriechu7@gmail.com}

\author[0000-0002-1437-4463]{Laurie E. U. Chu}
\altaffiliation{NASA Postdoctoral Program Fellow}
\affiliation{NASA Ames Research Center, Space Science and Astrobiology Division, Mail Stop 245-3, Moffett Field, CA 94035 USA}

\author{Klaus Hodapp}
\affiliation{Institute for Astronomy, 640 N. Auhoku Pl. \#209, Hilo, HI 96720 USA}


\begin{abstract}

Tracing dust in small dense molecular cores is a powerful tool to study the conditions required for ices to form during the pre-stellar phase.  To study these environments, five molecular cores were observed: three with ongoing low-mass star formation (B59, B335, and L483) and two starless collapsing cores (L63 and L694-2).  Deep images were taken in the infrared JHK bands with the United Kingdom Infrared Telescope (UKIRT) WFCAM (Wide Field Camera) instrument and IRAC channels 1 and 2 on the \textit{Spitzer Space Telescope}.  These five photometric bands were used to calculate extinction along the line of sight toward background stars.  After smoothing the data, we produced high spatial resolution extinction maps ($\sim$13-29\arcsec) .  The maps were then projected into the third dimension using the AVIATOR algorithm implementing the inverse Abel transform.  The volume densities of the total hydrogen were measured along lines of sight where ices (H$_2$O, CO, and CH$_3$OH) have previously been detected.  We find that lines of sight with pure \choh or a mixture of \choh with CO have maximum volume densities above 1.0$\times$10$^5$ cm$^{-3}$.  These densities are only reached within a small fraction of each of the cores ($\sim$0.3-2.1\%).  \choh presence may indicate the onset of complex organic molecule formation within dense cores and thus we can constrain the region where this onset can begin.  The maximum volume densities toward star-forming cores in our sample ($\sim$1.2--1.7$\times$10$^6$~cm$^{-3}$) are higher than those toward starless cores ($\sim$3.5--9.5$\times$10$^5$~cm$^{-3}$).




\end{abstract}



\section{Introduction} \label{sec:intro}

Dust grains and the ices that form on them play an important role in the early stages of stellar and planetary formation within dense molecular cores. The dense environments shield the innermost core from interstellar radiation. The extremely cold temperatures within the core allows gases to condense onto the surface of dust grains, adding layers and complexity to the icy surfaces they have gathered together.  

At relatively low densities \hto is the first to freeze out forming a monolayer of ice at an extinction A$_V$=1.6  \citep{Hollenbach2009}. As the density within the core increases, \hto is mixed with other ices such as CH$_4$, NH$_3$, and CO$_2$ and models show that at the highest densities (n$\geq$10$^5$ cm$^{-3}$) CO can completely freeze out \citep{Allamandola1999}.  At this point more complex ices can form such as H$_2$CO and CH$_3$OH on short time scales of $\sim$10$^4$ years \citep{Cuppen2009}. Models indicate these molecules are vital for the onset of complex organic molecules \citep{Fedoseev2017} and they appear to be abundant in cores even before stars form \citep{Chu2020, Scibelli2020}.

Ices have been observed along lines of sight through molecular cores toward background stars and Young Stellar Objects. Absorption features for \hto (3.0 $\mu$m), CO (4.67 $\mu$m) and CO$_2$ (4.27 $\mu$m) have been well characterized \citep[e.g.][]{Whittet1983, Whittet1985, Whittet1998,Chiar1995, Goto2018}.  Additionally CH$_3$OH (3.53 $\mu$m from the C-H stretching mode) has been detected along several lines of sight through molecular cores \citep[e.g.][]{Boogert2011, Chiar2011, Chu2020}.  These wavelengths are difficult to observe from the ground due to high sky background noise and telluric features.  Thus only a limited sample of background stars and YSOs have probed the environments for ice formation. Extinction thresholds have been determined for the formation of different ices within small isolated cores and large molecular clouds.  There are significant variations to this threshold between clouds, for example the extinction, \Av at which \hto ice forms in Ophichus is $\sim$10-15 magnitudes \citep{Tanaka1990} while the Pipe Nebula has a threshold A$_V$=5.2$\pm$6.1 mag \citep{Goto2018}.  Even smaller thresholds were found for the Taurus Molecular Cloud (TMC) and Lupus (A$_V$=3.2$\pm$0.1 and 2.1$\pm$0.6 mag respectively) \citep{Whittet2001, Boogert2013}. For CO ice the formation threshold is typically higher with A$_V$=6.0$\pm$4.1 for the TMC \citep{Chiar1995} and A$_V$=4.96$\pm$1.07 from isolated cores \citep{Chu2020}.  

Extinction thresholds can provide insight on the dust column density through the line of sight where ices can form.  However, mapping the extinction across the full core can provide spatial and structural context to understand what promotes or inhibits different ice growth.  Extinction maps have long been a tool to trace the dust in star forming regions.  Observations in the near-infrared (NIR) have utilized the fact that at longer wavelengths stars become less obscured by dust.  \citet{Lada1994} recognized this relationship of the infrared color excess with the extinction and used it to map dust in a method called NICE (Near-infrared Color Excess).  In this method the difference between the observed and intrinsic color of a background star can produce an extinction measurement.  The following equation  from \citet{Lada1994} is an example of a simplified way to calculate the extinction using photometry in the H and K bands assuming a normal reddening law from \citet{Rieke1985}:

\begin{equation}\label{eqn:extinction}
    (H-K)^{obs}=(H-K)^{int}+0.063 \times A_V.
\end{equation}

The "obs" represents the color that is observed while the "int" is the intrinsic color of the stars.  The H-K colors are beneficial to use because the intrinsic colors of stars have a small scatter \citep[e.g][]{Koornneef1983}.  Typically a nearby field of stars with low extinction levels is measured to obtain an average intrinsic color to use in the above equation.  After measuring the extinction along the line of sight toward background stars then the discrete measurements are smoothed into an extinction map.  This method has some limitations because using only one color can cause significant noise in the final \Av estimate.  Thus this method was later expanded into NICER where multiple wavelength bands could be used to determine the extinction with significantly smaller errors in \Av \citep{Lombardi2001}.
Additional improvements to this general technique have carefully removed some bias and inhomogeneities in the distribution of background stars that cause discrepancies on the small-scale structure \citep[NICEST,][]{Lombardi2009}.  New computational methods have also been employed for these techniques such as PNICER which uses a machine learning algorithm \citep{Meingast2017} and XNICER which uses a full Bayesian inference of the extinction for each observed object \citep{Lombardi2018}. 

Most recently, a new technique has been utilized to take two dimensional maps and project them into the third dimension using a mathematical technique called an inverse Abel transform.  The Abel transform has been used for example by taking a 3D object with optically thin emission and integrating the emission along the line of sight to produce a 2D image.  The \textit{inverse} Abel transform does the opposite where a 2D plane is projected into 3D using an axially symmetric distribution \citep{Abel1826}.  This technique has been used in several applications \citep[][and references therein]{Hasenberger2020}. In \citet{Hasenberger2020} they develop the algorithm AVIATOR to reconstruct 2D maps into 3D with assumptions that the distribution along the line of sight is similar to the distribution in the plane of projection.  Typically densities of molecular cores are derived from dust continuum emission maps in the submillimieter \citep[e.g.][]{Kirk2005, Ysard2012}.  However, they usually rely on an average temperature along the line of sight and do not consider temperature gradients within the cores.  Because these cores have stratified density structures that may have internal or external radiation fields, this can introduce errors on the density profile. Radiative transfer methods can also be used to model the volume density as shown in \citet{Nielbock2012, Steinacker2016} to incorporate temperature gradients but are model dependent. The inverse Abel transformation removes some uncertainties in these methods by not needing temperature information or model dependent parameters \citep{Hasenberger2020}.  Having three dimensional information is essential for observationally constraining the local densities of dust where ice forms.

In this work we present extinction maps and their three dimensional reconstructions for five cores at different evolutionary stages. All but one of these cores were previously studied for ongoing ice formation where column densities toward individual background stars were measured for H$_2$O, CO, and \choh ices \citep{Chu2020}.  This provides an excellent sample to compare the spatial densities of dust and gas required for the different ices to form.  We will first describe the molecular cores being investigated in Section \ref{sec:targets}.  Observations of the cores and data reduction methods are in Section \ref{sec:obs and red}.  In Section \ref{sec:results} we present the extinction maps of each core (Section \ref{sec:extinction maps}) and transform them into three dimensional maps (Section \ref{sec:Abel}) where spatial densities along lines of sight toward background stars with ice detections are shown.  We discuss these results in Section \ref{sec:discussion}.

\section{Target Selection} \label{sec:targets}

We have selected five molecular cores for extinction mapping and analyzing the local densities where ices form.  These are all small ($\sim$0.2-1 pc) dense mostly isolated cores.  They represent different stages of evolution where two are collapsing (L63 and L694-2), two have Class 0 Young Stellar Objects (YSOs) embedded in the core (B335 and L483), and one is quiescent with several later stage (Class II) YSOs (B59). All of the cores were chosen because they have a high density of background stars with a position against the galactic bulge.  This allows for measurements of many lines of sight through the cores so that the extinction maps will have very fine sampling with high spatial resolutions on the sky.  Because the cores are also nearby ($\leq$250 pc) there are very few foreground stars contaminating the data. The galactic bulge contains mostly more evolved late K and M stars which have a small spread in color reducing the uncertainty in the intrinsic color measurements used in Equation \ref{eqn:extinction} \citep{Zoccali2003}.  As previously mentioned, four of the five cores were studied in \citet{Chu2020} where several lines of sight displayed the presence of H$_2$O, CO, and CH$_3$OH ice.  The three dimensional maps of the cores will allow us to probe the spatial density of hydrogen required for this ice formation.  The cores are relatively simple in shape and structure which is not necessarily the most representative of other star forming regions, but these properties will help simplify the three dimensional reconstruction. Each of the cores are summarized in Table \ref{tab:core_properties} and below we highlight some of the features seen in the infrared and the ices that have been detected.

\begin{table*}[h!t]
\tabcolsep=0.1cm
\tabletypesize{\small}
\centering
\caption{Molecular Core Sample}
\label{tab:core_properties}
\begin{tabular}{lcccccl}

\tableline\tableline

Core	&	RA	&	Dec	&	Distance (pc)\footnote{Distances to the core with references and uncertainties cited in the text}	&	Size (arcmin)\footnote{Size in the RA direction of the core as shown in Figure \ref{fig:colorext}}	&	Size (pc)\footnote{Size of the core in the RA direction shown in Figure \ref{fig:colorext} for the given distance quoted. Sizes were determined by visually examining the JHK color composite images to include the most reddened sources in the core and any extended features protruding from the central core.}	&	Evolutionary Stage	\\
\tableline
L63	&	16:50:14.9	&	-18:06:23	&	130	&	13.18	&	0.50	&	Collapsing, starless	\\
B59	&	17:11:21.6	&	-27:27:42	&	163	&	16.72	&	0.79	&	Stable, star-forming	\\
L483	&	18:17:29.8	&	-04:39:38	&	225\footnote{Average taken for 200-250 pc}	&	8.07	&	0.52	&	Class 0 star-forming	\\
B335	&	19:37:01.0	&	+07:34:10	&	150	&	3.77	&	0.16	&	Class 0 Star-forming	\\
L694-2	&	19:41:04.5	&	+10:57:02	&	230	&	9.95	&	0.67	&	Collapsing, starless	\\

\tableline

\vspace{0.5cm}
\end{tabular}
\end{table*}

\subsection{L63}

L63 is one of the densest regions in the northern Ophiuchus (Oph N) complex and is classified as being starless in a quasi-equilibrium collapsing stage \citep[e.g][]{Nozawa1991, WardThompson1994, WardThompson1999, Kirk2005, Seo2013}. The distance to Oph N has been estimated from Hipparcos parallaxes to be 145 $\pm$2 pc \citep{deGeus1989} but extinction-based distance modulus estimates have a range of distances between 80-200 pc where some individual clouds  are further away than others \citep{deGeus1989, Straizys1984, Wilking2008}.  In \citet{Hatchell2012} an average distance of 130 pc is used and we adopt the same for the distance to L63.  The density of background stars for L63 is lower than others in our sample and there were not any background stars that were bright enough to be included in the study by \citet{Chu2020} for the detection of ices.  

\subsection{B59}

B59 is the densest and only known part of the Pipe Nebula where star formation has begun \citep{Onishi1999, Forbrich2009}.  Using astrometric data from Gaia, the distance is measured with high precision at 163$\pm$5 pc \citep{Dzib2018}.  It is the largest core in our sample with the densest regions covering $\sim$0.8$\times$0.8 pc and a total mass of $\sim$30 M$_{\odot}$ \citep{DuarteCabral2012}.  \citet{Brooke2007} developed an extinction map for B59 with 2MASS JHK data with a spatial resolution of $\sim$100$^{\prime\prime}$.  They found a peak extinction of \Av$\sim$45.  Within $\sim$0.1 pc of the highest density peak 13 low-mass YSOs were identified \citep{Brooke2007}.  Most are classified as later Class II sources but two are Class 0/I or I. In \citet{Chu2020} spectra of five of the Class II YSOs were observed where \hto ice was detected along the line of sight to all five, four displayed frozen CO of which two had a mixture of CO with a polar ice, presumably CH$_3$OH.  They determine that the ices are most likely present in the foreground cloud and not part of the YSO's disk or envelope.

\subsection{L483}

L483 is a small isolated core with an embedded low mass \citep[0.1-0.2\Msun,][]{Oya2017} protostar, IRAS 18148-0440 \citep{Parker1988}.  Though the protostar is deeply embedded at \Av$\sim$70, it appears to be the source of a variable nebula that is only visible in the infrared \citep{Connelley2009}.  The variability is most likely due to opaque clouds within $\sim$1 AU that cast shadows and change the illumination of the nebula. In \citet{Fuller2000} they model the central region within 3000 AU of the source to have infalling material from the surrounding core.  L483 is typically associated with the Aquila Rift region at a distance of 200 pc \citep{Dame1985} but VLBA and Gaia-DR2 astronmetry show that the distance to Aquila is upward of 436$\pm$9 pc \citep{Ortizleon2018} but the parallaxes and extinction of stars near L483 still indicate a closer distance of 200-250 pc \citep{Jacobsen2019} and thus we adopt the average distance of 225 pc. Signatures of \hto, CO$_2$, CO and \choh ice have been measured through lines of sight through this core to background stars outside of the influence of the protostellar envelope \citep{Boogert2011, Chu2020}.

\subsection{B335}

B335 is a Bok globule  with a Class 0 YSO deeply embedded in the core \citep{Frerking1982, Keene1983, Chandler1990, Hodapp1998}.  The protostar appears to be less evolved than the protostar in L483 with a lower mass \citep[0.02-0.06 \Msun,][]{Imai2019} and very low levels of rotation \citep{Jacobsen2019}.  In the infrared there are emission features where a bipolar outflow from the protostar has broken through the globule on both sides.  At 8 $\mu$m a shadow extends $\sim$3000-7500 AU with \Av$>$100 and core shine can be seen at shorter wavelengths \citep{Stutz2008}.  The distance is 150$^{+50}_{-90}$ pc but with higher uncertainty than some of the other cores in our sample \citep{Stutz2008}. There is a very bright background star (7.6 mag in the K band) near the center of the core (2MASS 19365867+0733595).  It has been studied in \citet{Chu2020} and along the line of sight H$_2$O and CO ices are detected.  They also find CO mixed with polar ice (most likely \choh ice) but the upper limits for detecting \choh independently are very low.  This is most likely because at a relatively low extinction of \Ak$=1.29\pm0.05$ the line of sight does not sample a high enough density for an abundance of \choh ice to form.   

\subsection{L694-2}

The isolated core L694-2 is collapsing with a strong infall with speeds faster than those within L63 and increasing toward the center \citep{Lee2004, Seo2013}. However, the core appears to be starless \citep{Harvey2002, Harvey2003, Evans2003, Suresh2016}.  There is a compact centrally condensed core \citep{Lee2001} with a mass of $\sim$1\Msun and within the next 10$^4$ years a point mass is expected to form \citep{Williams2006}. The distance to L694-2 was originally assumed to be similar to the distance of B335 since it is nearby \citep{Tomita1979} but \citet{Kawamura2001} revised this distance based on star counts to be 230$\pm$30 pc which is adopted here.  In \citet{Chu2020} ice features were detected toward background stars including H$_2$O and CO. For one line of sight just outside the densest part of the core \choh ice was detected.    

\section{Observations and Data Reduction} \label{sec:obs and red}

\subsection{UKIRT Photometry}

Observations of each core were taken in the J (1.25 $\mu$m), H (1.64 $\mu$m), and K (2.20 $\mu$m) band filters \citep{Hewett2006} with the 3.8m United Kingdom Infrared Telescope (UKIRT) using the Wide Field Camera (WFCAM) instrument \citep{Casali2007}. WFCAM has four 2048$\times$2048 pixel detectors each with 0\farcs4 resolution for a total area of coverage of 0.19 deg$^2$ per detector.  Integration times were typically longer for the J band than the H and K bands and ranged from 1.0-5.6 hours (total integration times for each band are presented in Figure \ref{fig:colorext}). Observations were made during the 2014A and 2016A semesters.    

Images were processed with the Cambridge Astronomical Survey Unit as described in \citet{Irwin2004} and retrieved from the WFCAM Science Archive \citep{Hambly2008}.  All of the cores fit onto one detector except for B59 where we treated each chip individually for analysis before combining as a mosaic image.  Using stacked images we obtained the positions of stars using IRAF's \texttt{daofind} with a FWHM of 5 pixels and a 4$\sigma$ detection threshold. Then using this source list we extracted the stellar photometry in each band for individual frames using PSF fitting with the IRAF routine \texttt{daophot}.  A sample of 18 stars for each core was used to produce a model PSF that was used to obtain the photometry for all stars.  Aperture photometry was not possible because of the crowded star fields.  To calibrate the photometry we used the source catalogs that are also products of the standard reduction pipeline \citep{Hodgkin2009} that are calibrated to the Two-Micron All-Sky Survey \citep[2MASS][]{Skrutskie2006}. We matched these catalogs from each individual frame to find the average photometry for each star.  Then using this master list any stars in the 12-18 magnitude range with errors $\leq$0.07 mag were matched to targets we detected and a calibration was done for each frame.  The final photometry was an average between the individual frames.  The limiting magnitudes vary slightly between the different cores depending on the total integration times and are complete to the 90\% level corresponding to $\sim$21.0 mag in J band and $\sim$20.0 mag in H and K bands. Stars were saturated below $\sim$10 mag.

\subsection{Spitzer Photometry}

Our five cores were observed with \textit{Spitzer} IRAC's channel 1 and 2 \citep[3.6 and 4.5 $\mu$m respectively,][]{Fazio2004} during the warm mission (cycles 11 and 12, program ID 11028).  Observations had 100 second exposures with total integration times of one hour per core producing a 1$\sigma$ limiting flux of 0.36 $\mu$Jy in Channel 2.  This was significantly deeper than previous observations of the same fields found in the \textit{Spitzer} Heritage Archive with integrations of only $\sim$100 seconds producing 1$\sigma$ limiting fluxes of 2.16 $\mu$Jy (Program ID 139, PI Evans, N. for B59, L63, L483, and L694-2; Program ID 94, PI Lawrence, C. for B335; Program ID 20119, PI Lada, C. for B59). Observations for our program were made in the HDR mode and channel 2 observations were taken immediately after channel 1 in order to remove any possible uncertainty due to variability in the stars.  The dither pattern used a 36 point dither.

Images were processed by the \textit{Spitzer} Science Center pipeline S19.2.0. The photometry was measured using mosaics created with \texttt{MOPEX} following the guidelines and best inputs explained in the IRAC Instrument Handbook v2.1.  We performed Point Response Function (PRF) photometry with the \texttt{APEX} procedure because PSF fitting with IRAC has proven to be problematic in channels 1 and 2. 

\section{Results} \label{sec:results}

\subsection{Extinction Mapping} \label{sec:extinction maps}

Employing the NICER method from \citet{Lombardi2001} we derive extinction maps for the five cores using multiple photometric bands.  Equation \ref{eqn:extinction} shows a simple method to determine the extinction using only the H-K color but NICER can implement multiple color measurements to reduce errors in the final extinction measurement. First we use the JHK data and \textit{Spitzer} IRAC Channels 1 and 2 (ch1 and ch2) to take the covariance matrix of the J-H, H-K, K-ch1, and ch1-ch2 intrinsic colors.   The reddening vectors (A$_{\lambda}$/A$_V$) for the different color combinations are taken from \citet{Rieke1985} for JHK colors and from \citet{Indebetouw2005} for IRAC channels.  In \citet{Indebetouw2005} the extinction law is derived using A$_K$ as a reference rather than A$_V$ and can vary between different environments.  The variation is characterized by the reddening parameter $R_V=A_V/E_{B-V}$ where $E_{B-V}$ is the color excess. For $R_V$=3.1, $A_V/A_K\sim8.8$ but for regions where $R_V$=5, $A_V/A_K\sim7.5$ \citep{Cardelli1989}.  We adopt $R_V=3.1$ for all of the coefficients used in the NICER equations since we have used the color combinations from \citet{Rieke1985} for JHK, which is appropriate for the $R_V=3.1$ environment.  Knowing the reddening vectors and intrinsic colors then allows us to calculate the \Av along the line of sight toward each background star. Following are the NICER equations used similar to that found in Equation \ref{eqn:extinction}:
\begin{equation}\label{eqn:JH}
    (J-H)^{obs}=(J-H)^{int}+0.11 \times A_V
\end{equation}
\begin{equation}\label{eqn:HK}
    (H-K)^{obs}=(H-K)^{int}+0.063 \times A_V
\end{equation}
\begin{equation}\label{eqn:Kch1}    
    (K-[3.6])^{obs}=(K-[3.6])^{int}+0.048 \times A_V
    \end{equation}
\begin{equation}\label{eqn:ch1ch2}    
    ([3.6]-[4.5])^{obs}=([3.6]-[4.5])^{int}+0.015 \times A_V.
\end{equation}

We do not use the NICEST routine \citep{Lombardi2009} since our cores are relatively simple in structure and nearby to where contamination by foreground stars is not significant.  Even with some foreground stars, the density of background stars is high and the foreground stars would appear as outliers and have a small impact on the final extinction map.

\subsubsection{Intrinsic Colors}

In previous studies \citep[e.g.][]{Lada1994, Lombardi2001, Lombardi2006, Lada2009} average intrinsic colors of background stars are obtained by using a nearby control field with negligible extinction levels. This also accounts for any reddening due to intervening dust in the interstellar medium (ISM).  However, in our study we did not obtain nearby control fields for each core since it would have nearly doubled our observing time.  Thus we demonstrate a two-step approach to simulate a control field.

Using galactic simulations of stars from the TRILEGAL model\footnote{http://stev.oapd.inaf.it/cgi-bin/trilegal} \citep{Girardi2005} we can estimate the intrinsic colors of background stars in the direction of each molecular core individually.  This model uses four major components to simulate stars in a particular region of the sky: 1. A library of stellar evolutionary tracks, 2. A library of synthetic spectra, 3. The instrumental setup for a given telescope, and 4. A description of Galaxy components such as the Galactic thin
and thick disk, Halo, and Bulge. The input requires the equatorial or galactic inputs for the region of the sky of interest and a total field area dimension. We chose the coordinates that are close to the center of each core with an area of 0.1 deg$^2$. Then a photometric system is selected with a chosen magnitude limit where we set the limiting magnitude to 20 mag in H band. Other parameters such as the IMF, binary fraction, and galactic components were kept at the default and the dust extinction was set to zero.  The output produces a list of stars with expected stellar parameters for the particular region of the sky including temperature, mass, and the expected magnitude in the 2MASS JHK and \textit{Spitzer} IRAC channels 1-4 filters.  The distribution from the simulations allow us to identify the expected intrinsic color for the four color combinations mentioned above (J-H, H-K, K-ch1, and ch1-ch2). 

To ensure that the simulation produced accurate colors we tested a field of stars used in \citet{Lada1994} - IC 5146 (RA: 328.4577, Dec: 47.2566 (J2000)).  They determine the average H-K color to be 0.13 but in our simulation with a field of view of 0.5 deg$^2$ and a limiting magnitude of 14 in K band (corresponding to the approximate limiting magnitude in their work) we find an intrinsic H-K color of 0.063.  This is because the theoretical values do not account for any interstellar dust in the foreground or behind the molecular core, along the line of sight.  Using the Galactic Dust Reddening and Extinction website\footnote{https://irsa.ipac.caltech.edu/applications/DUST/}, the data from the Infrared Astronomy Satellite (IRAS) mission and the DIRBE experiment (Diffuse InfraRed Background Experiment) onboard the COBE satellite produce the visual extinction for a given location \citep{Schlegel1998}.  Newer estimates of dust reddening were also measured by \citet{Schlafly2011} using the Sloan Digital Sky Survey.  Toward a low extinction region near IC 5146 (RA: 328.4-329.0, Dec: 47.5-47.7) the \Av is $\sim$0.72-0.87 (depending on which reddening data is used).  Using the average \Av$=0.80$ and the reddening law used in \citet{Rieke1985} (to remain consistent with the NICER calculations), the ISM reddening of H-K is $\sim$0.05 and thus the sum of the intrinsic color and ISM reddening in H-K is $\sim$0.113, which is in agreement with \citet{Lada1994}. 

For each of the cores in our sample we calculate the sum of the intrinsic and ISM colors in this way to determine the overall extinction.  Errors on the intrinsic colors come from the standard deviation of the intrinsic colors from the TRILEGAL simulation.  Table \ref{tab:ext parameters} reports these values.  The errors for E(B-V) are provided by the Galactic Dust Reddening and Extinction website and we propagate the errors from both reddening estimates to determine the error on the ISM extinction (Table \ref{tab:ext parameters}).  The ISM extinction can introduce systematic errors to the overall extinction measurement depending on the extinction law adopted. Since we use the relation in \citet{Rieke1985} for the NICER technique, we use the ISM A$_V$ where R$_V=3.1$ throughout this work to remain consistent, but dense cores may have a higher reddening vector such as R$_V=5.5$ \citep{Weingartner2001}.   We explore how this could affect the overall extinction and in Table \ref{tab:ext parameters} we show the average \Av for the ISM used for each core with both R$_V=3.1$ and R$_V=5.5$.  The difference between the two extinction laws introduces a discrepancy in A$_V$ by $\sim$1 magnitude (or $\sim2$ magnitudes for L483).  In low extinction regions this could have a large effect, but in higher extinction regions a difference in the ISM of $\sim$1-2 magnitudes becomes small.  However, if the extinction in the ISM may increase by a factor of $\sim$1.8 depending on whether the R$_V$ chosen is 3.1 or 5.5, this suggests the A$_V$ in the core could potentially be underestimated by a similar factor when using R$_V=3.1$.  Until we have a better understanding of the extinction law in dense cores, we are limited to this additional systematic error for the A$_V$ in the core.   

\begin{table*}[h!t]
\tabcolsep=0.2cm
\tabletypesize{\small}
\centering
\caption{Extinction Mapping Parameters}
\label{tab:ext parameters}
\begin{tabular}{lccccccc}

\tableline\tableline

Core	&	J-H\footnote{\label{note1}Intrinsic Colors from TRILEGAL Simulations including the intervening interstellar dust extinction.  Errors shown are the standard deviation of the intrinsic colors.}	&	H-K\footref{note1}	&	K-Ch1\footref{note1}	&	Ch1-Ch2\footref{note1}\footnote{\textit{Spitzer} IRAC channels 1 and 2}	&	ISM$_R$ A$_V$\footnote{The interstellar dust extinction from the Galactic Dust Reddening and Extinction website using the extinction law from \citet{Rieke1985} with reddening R$_V$=3.1}	& ISM$_W$ A$_V$\footnote{The interstellar dust extinction from the Galactic Dust Reddening and Extinction website using the extinction law from \citet{Weingartner2001} with reddening R$_V$=5.5} & Smoothing (\arcsec)\footnote{Smoothing resolution used for extinction maps}\\
\tableline
L63	&	0.616$\pm$0.120	&	0.186$\pm$0.068	&	0.162$\pm$0.096	&	0.024$\pm$0.038	&	0.98$\pm$0.10	& 1.77$\pm$0.18 &	29.2	\\
B59	&	0.440$\pm$0.107	&	0.109$\pm$0.022	&	0.067$\pm$0.013	&	-0.015$\pm$0.025	&	0.99$\pm$0.09	& 1.75$\pm$0.16 &	13.6	\\
L483	&	0.804$\pm$0.122	&	0.299$\pm$0.071	&	0.258$\pm$0.109	&	0.050$\pm$0.040	&	2.69$\pm$0.25	& 4.75$\pm$0.45 &	13.0	\\
B335	&	0.530$\pm$0.146	&	0.130$\pm$0.064	&	0.092$\pm$0.084	&	-0.001$\pm$0.040	&	0.73$\pm$0.14	& 1.31$\pm$0.12 &	17.0	\\
L694-2	&	0.613$\pm$0.146	&	0.174$\pm$0.063	&	0.136$\pm$0.081	&	0.010$\pm$0.039	&	1.10$\pm$0.15 & 2.03$\pm$0.26	&	16.9	\\

\tableline

\vspace{0.5cm}
\end{tabular}
\end{table*}

\subsubsection{Map Smoothing}

Once the extinction toward each line of sight through the core is calculated, extinction maps are constructed.  L483, B335 and B59 harbor YSOs but the YSOs were excluded in producing the extinction maps.  The YSOs in L483 and B335 are highly extincted and thus were not even detected in the \textit{Spitzer} data.  For B59 we also removed 13 YSOs in our sample that were identified by \citet{Forbrich2009} and \citet{Brooke2007}.  These YSOs include the five identified in Section \ref{sec:lines of sight}; the remaining seven are identified in Figure \ref{fig:colorext}. The extinction maps are then created using a weighted mean smoothing with a Gaussian smoothing kernel.  The weighted mean assumes that all of the stars are in fact background stars and because our cores are very nearby, this assumption is appropriate and does not introduce significant bias.  To determine the smoothing parameter we use the median distance to the tenth nearest neighbor of each star.  This means in the highest extincted regions the resolution is undersampling the data but we find the resolution to be a good balance in order to not lose actual structure within the core. The grey-scale smoothed extinction maps are shown in Figure \ref{fig:colorext} beside color composite images of the cores. Contours are shown representing levels with $\Delta$A$_V$=4. The smoothing resolutions are listed in Table \ref{tab:ext parameters} for each core and displayed on the extinction maps.  

In order to demonstrate where the extinction measurement is unreliable due to undersampling, regions in the center of each core are shown in black where the third nearest neighbor is less than twice that of the smoothing resolution.  This means the extinction value for the given pixel is interpolated from three or fewer nearby stars. For L63, there were no regions that were undersampled by this criteria.  Additionally, the errors on the A$_V$ value for each pixel were determined as the interpolated value of the variance divided by the effective number of stars that contribute to the measurement of that pixel.  The variance is converted to a 1$\sigma$ value and we divide the 1$\sigma$ values by the A$_V$ for each pixel to determine the error fraction.  We then determine the median error fraction for each core in the range $4 \leq A_V \leq 32$ and $A_V > 32$ to represent the significance of the errors and how the errors may be impacted at higher levels of extinction.  We do not consider low extinction regions ($A_V < 4$) since, as discussed previously, the errors on the intrinsic colors and extinction law can be rather significant and dominate.  In table \ref{tab:Av err} we convert this error fraction to a percent and find that for extinction values $4 \leq A_V \leq 32$ the errors are on the 2.5-8.2\% level.  For higher extinction regions the errors are slightly worse except for L63 and B59 where the high extinction regions are still decently sampled.   

\begin{table*}[h!t]
\tabcolsep=0.1cm
\tabletypesize{\small}
\centering
\caption{Median A$_V$ Errors}
\label{tab:Av err}
\begin{tabular}{l|c|c}

\tableline\tableline

Core & 4 $\leq$ A$_V$ $\leq$ 32\footnote{The median errors calculated as $\sigma_{A_V}$/A$_V$ and converted to a percentage for all pixels with 4 $\leq$ A$_V$ $\leq$ 32}  & A$_V > 32$\footnote{The median errors calculated as $\sigma_{A_V}$/A$_V$ and converted to a percentage for all pixels with A$_V$ $>$ 32} \\
\tableline
L63 & 5.7\% & 4.6\% \\
B59 & 2.5\% & 0.8\% \\
L483 & 6.1\% & 8.8\% \\
L694-2 & 8.2\% & 10.4\% \\
B335 & 5.0\% & 5.9\% \\

\tableline\tableline

\end{tabular}
\end{table*}

\begin{figure}[!htb]
\centering
\includegraphics[scale=0.55]{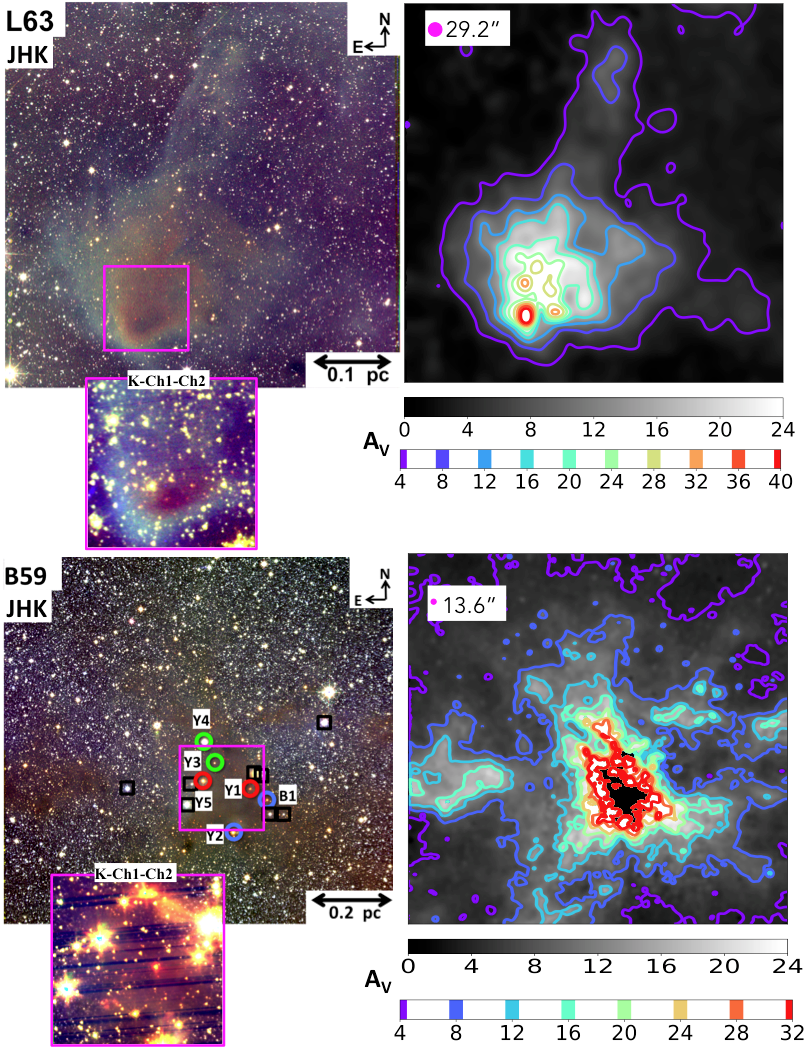}
\caption{On the left JHK color composite images of each core are shown with the size scale labeled using the reported distances to the cores in the text.  The following are the total exposure times used for each cloud: \textit{B59}: J=1.0 hr, H=1.2 hr, K=1.4 hr, \textit{L483}: J=5.6 hr, H=2.8 hr, K=2.8 hr, \textit{B335}: J=4.0 hr, H=2.8 hr, K=3.2 hr. \textit{L694-2}: J=3.2 hr, H=2.0 hr, K=2.0 hr. \textit{L63}: J=4.0 hr, H=3.2 hr, K=3.6 hr. On some of the images circles show lines of sight that have ice measurements from \citet{Chu2020} where blue has \hto ice, green has \hto and CO ice, red has \hto, CO, and CO$_P$ ice, and yellow has \hto, CO, CO$_P$, and CH$_3$OH ice (CO$_P$ is CO mixed with polar ice).  Black squares identify other YSOs where lines of sight for ices were not measured.  The magenta box in the JHK color images is also shown in color images with K, Channel 1 and 2 from \textit{Spitzer} to display background stars detected in the highest density region. The right panels are the corresponding grey-scale extinction maps where the color bar is in A$_V$. The black centers of some cores show where the extinction level saturates.  Overlaid contours have step sizes A$_V$=4 represented by different colors.  The size of the smoothing resolution is shown in the upper left.}
\label{fig:colorext}
\label{fig:L63_B59ColorExt}
\end{figure}

\begin{figure}[!htb]
\ContinuedFloat\centering
\includegraphics[width=1\textwidth]{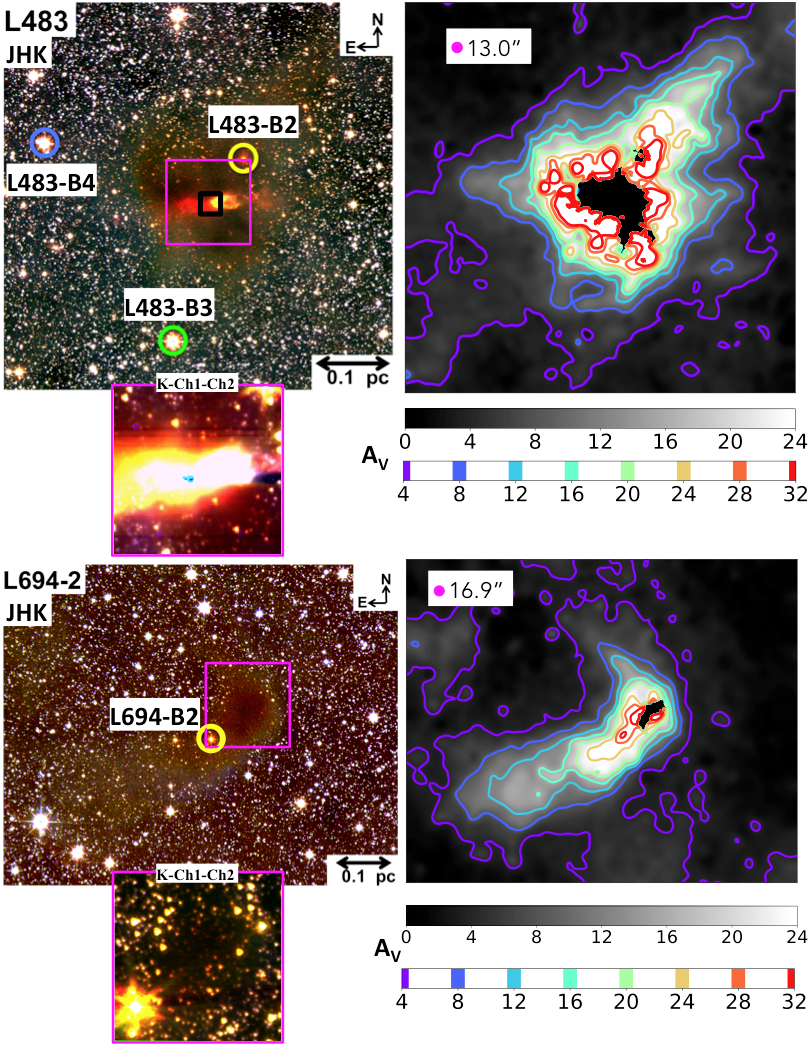}
\caption{(Continued)}
\label{fig:L483_L694ColorExt}
\end{figure}

\begin{figure}[!htb]
\ContinuedFloat\centering
\includegraphics[width=1\textwidth]{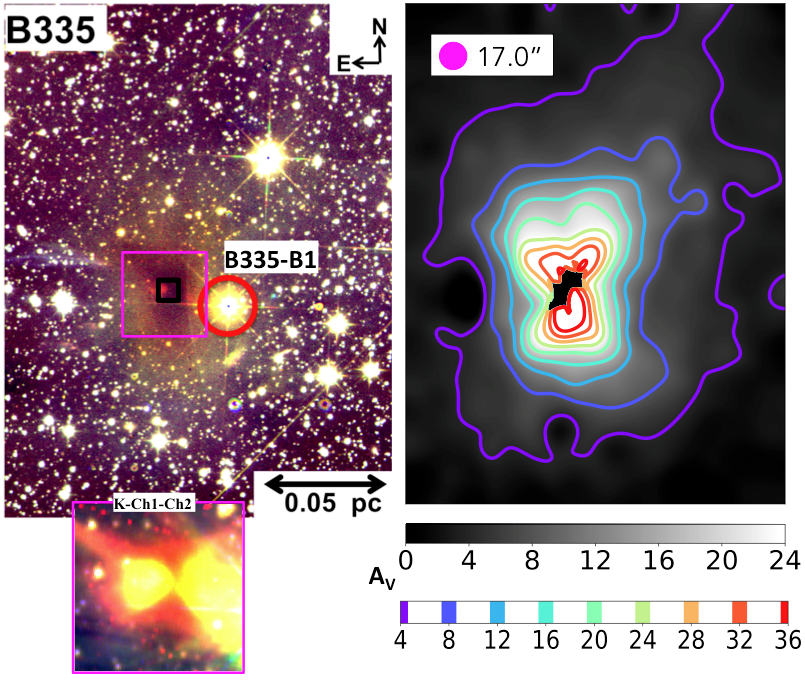}
\caption{(Continued)}
\label{fig:B335ColorExt}
\end{figure}

\subsection{Three Dimensional Reconstruction of Molecular Cores} \label{sec:Abel}

Each of the five extinction maps were prepared for the AVIATOR algorithm to reconstruct a three dimensional volume density distribution from the two dimensional maps \citep{Hasenberger2020}.  This algorithm calculates the volume density assuming a distribution perpendicular to the plane of projection is similar to that in the plane of projection, without requiring symmetry in the plane of projection.  Because of computational time restrictions the maps were first binned to have a resolution of 2.5~\arcsec/pixel (B335), 4~\arcsec/pixel (L483,) 5~\arcsec/pixel (L694-2), 6~\arcsec/pixel (L63), and 9~\arcsec/pixel (B59).  These pixel sizes are smaller than the beam sizes used for smoothing the extinction maps and this causes oversampling meaning that the pixel sizes do not necessarily represent the resolutions in the 3D maps.  Threshold levels of the extinction are required for the reconstruction and we chose a threshold with a stepsize of 0.02 meaning that each threshold level is at least 2\% higher than the previous level.  The algorithm then produces a 3D data cube (map) of extinction where the sum along the lines of sight reproduces the original two-dimensional map with errors typically on the 2-5\% level and increasing to errors of A$_V=2$ in the highest extincted regions highlighted as undersampled regions in Figure \ref{fig:L63_B59ColorExt}. The undersampled regions were used in creating the 3D maps as they retain information related to the ice chemistry in the densest parts of the core (see Section \ref{sec:lines of sight} for further discussion.)

\begin{figure}[!htb]
\centering
\includegraphics[width=1\textwidth]{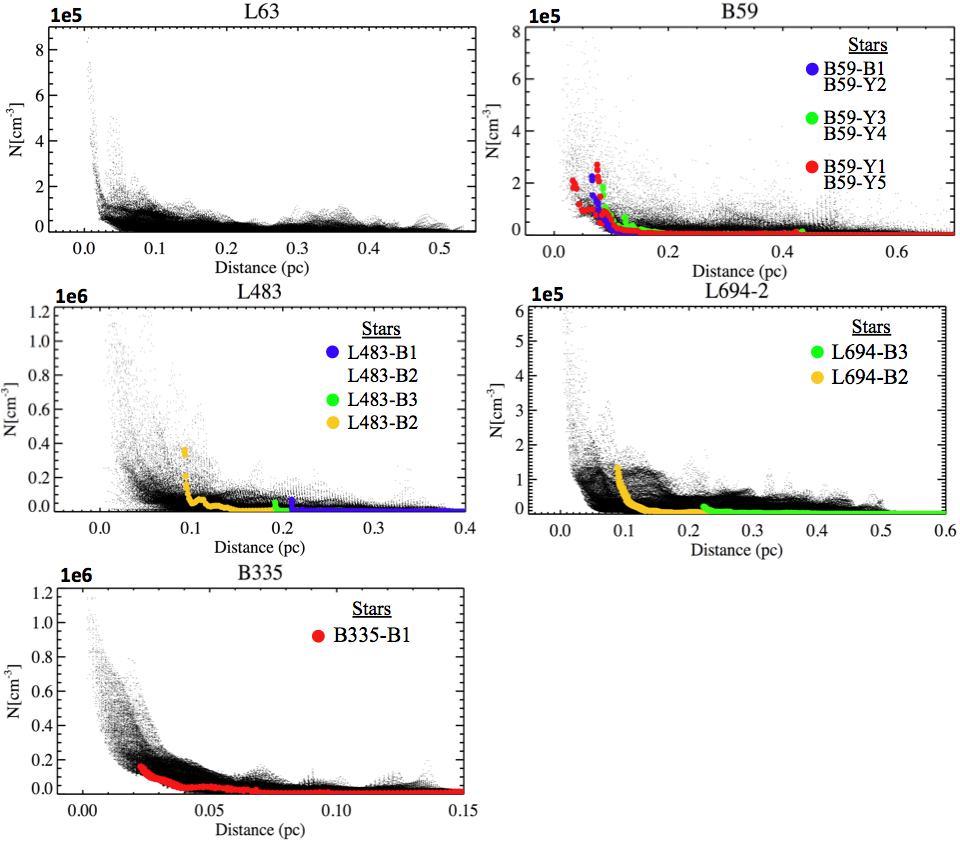}
\caption{The voxel volume density of each point within the core as a function of distance from the densest point in the core derived from the AVIATOR algorithm (Black points).  The colored points show the specific lines of sight toward YSOs and background stars. The colored data are only showing the lines of sight through the pixel corresponding to the center of the star's right ascension and declination. The colors are the same as Figure \ref{fig:colorext} representing the different ices detected.}
\label{fig:radprofile}
\end{figure}

For each core we find the point with the maximum extinction within the core.  Because the dimension of each voxel (three dimensional pixel) is the same, the size of a single pixel in the x and y direction (assuming the distances to the cores are accurate in Table \ref{tab:core_properties}) is the same as along the z direction.  This allows us to calculate the distance for every voxel in the core from the point with the maximum extinction. Using this distance a volume density of hydrogen (n[cm$^{-3}$]) can also be determined where the relation N$_H$/A$_V$ = 1.79$\times10^{21}$ cm$^{-2}$ mag$^{-1}$ represents the total hydrogen column (H and H$_2$) \citep{Predehl1995}.  This relationship is discussed further in Section \ref{sec:Nh Av}.  Figure \ref{fig:radprofile} shows the volume density (n[cm$^{-3}$]) for each voxel as a function of distance from the maximum density in parsecs.  The voxel with the maximum density in each core comes from more undersampled regions in the core, but since the undersampled regions are due to high extinction representing the center of the core, we can still use it to draw conclusions about the core structures and make comparisons between the cores.  Errors on the maximum volume density are calculated as the standard deviation of the density values for the surrounding voxels out to the distance of the smoothing resolution for each core (Table \ref{tab:lines of sight}).  In some cases this produces errors that are unreasonably small as the voxels are not completely independent values, which is required when taking the standard deviation. These uncertainties also do not account for errors arising from the AVIATOR reconstruction, the variation in the N$_H$/A$_V$ relationship, and the uncertainties in distances to the cores, which are all discussed separately, and in some cases are much larger than those found from the standard deviation.   Below are descriptions of the features observed in these figures and some analysis with comparisons to features seen in the extinction maps:

\textbf{\emph{L63 - }}  The highest density region appears to be a very small region within the core and there is a steep dropoff to lower density regions.  There is a second peak in the density at a little further distance ($\sim$0.05 pc) from the highest density region which is another dense region that can be seen in the extinction map.  In the 2D map we see a peak in the southeast portion of the cloud and two other dense regions north and west of the highest density region.  These dense regions are possibly artifacts from the smoothing kernel used because the structures are smaller than the smoothing resolution. The maximum density reached is (9.0 $\pm$ 1.7)$\times$10$^5$ cm$^{-3}$. 

\textbf{\emph{B59 - }} This core has a very broad region of high density extending to distances beyond 0.1 pc.  The peak density is the highest of any of the cores and is not shown in the figure at (1.7$\pm$0.5)$\times$10$^6$ cm$^{-3}$.  There are several high density peaks  ($>$2$\times$10$^5$ cm$^{-3}$) at much further distances.  The extinction map also shows a very irregular shape with spots of higher extinction.  This core is actively forming several stars and it is evident that there are multiple areas with higher densities for this star formation to occur.

\textbf{\emph{L483 -}} Similar to B59, the relationship is somewhat broad at the highest density levels.  It then has several parts of the core that reach a density between $\sim$3$\times$10$^5$ cm$^{-3}$ and $\sim$6$\times$10$^5$ cm$^{-3}$ and this represents the more extended feature in the northwest direction from the core (Figure \ref{fig:L483_L694ColorExt}).  This core has one of the highest spatial densities out of the five cores at (1.5$\pm$0.5)$\times$10$^6$ cm$^{-3}$.  However, it also has one of the highest uncertainties in distance estimates and the standard deviation is high, making this maximum density more unreliable.

\textbf{\emph{L694-2 - }}  There is a high peak in the density reaching (6.5$\pm$1.3)$\times$10$^5$ cm$^{-3}$ that drops off somewhat rapidly to lower density regions.  There are many points in the core that reach a density $\sim$1.5$\times$10$^5$ cm$^{-3}$ that extends nearly 0.2 pc away from the densest region.  Similar to L483, this is most likely due to the elongated feature of infalling material in the southeast direction in the core (Figure \ref{fig:L483_L694ColorExt}).

\textbf{\emph{B335 - }} This core appears in the extinction map to be the most spherical and the data in Figure \ref{fig:radprofile} reflects that.  It has a somewhat smooth trend from high to low densities as a function of distance from the highest density region.  The volume density is also comparable to the other star-forming cores reaching a maximum of (1.2$\pm$0.2)$\times$10$^6$ cm$^{-3}$.

\subsubsection{Lines of Sight with Ices} \label{sec:lines of sight}

Knowing the extinction distribution along lines of sight through the cores and thus the spatial density, we can compare lines of sight where various ices are detected.  From \citet{Chu2020} there are 13 lines of sight through four of the cores that have a 3$\sigma$ detection of ice where eight are background stars and five are YSOs (Class II from the B59 core).  Four sources have only \hto ice, four have \hto and CO ice, three have \hto, CO, and a mixture of CO with polar ice (CO$_P$), and two have all of these ices and additionally \choh ice.  The ice column densities for each are in Table \ref{tab:lines of sight}, and Figure \ref{fig:L63_B59ColorExt} identifies where the ices are located in the core.  

\begin{figure}[!htb]
\centering
\includegraphics[scale=0.2]{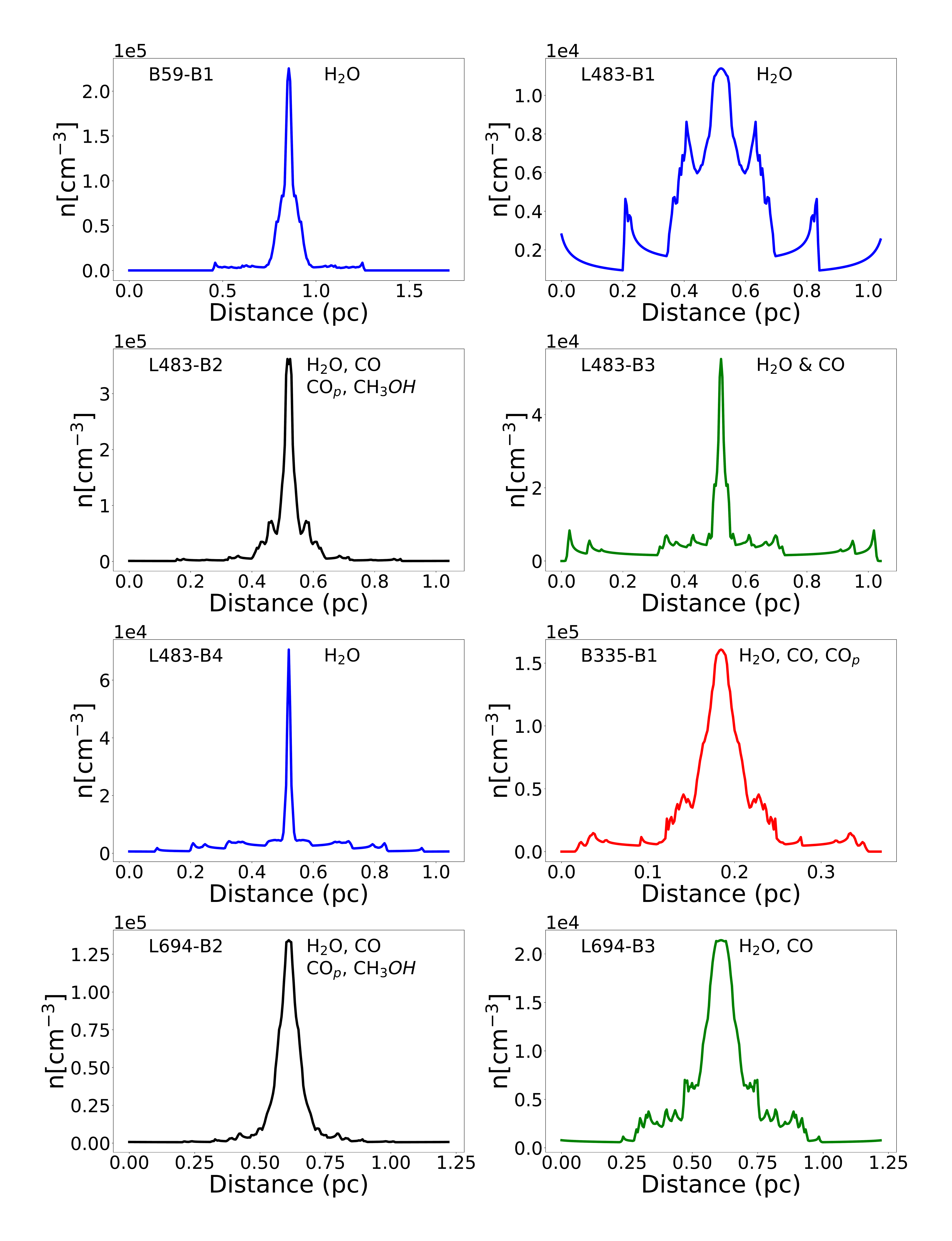}
\caption{The volume density distribution of hydrogen (H and H$_2$) along the line of sight toward background stars with different ices detected is shown.  The volume densities are calculated using the relationship between N$_H$ and A$_V$ explained in Section \ref{sec:Abel}. The distances on the x-axis are calculated based on the distances to the cores in Table \ref{tab:core_properties}.  The blue data represents lines of sight with only H$_2$O ice detected; green data shows where \hto and CO ice are detected; red data is for lines of sight with \hto, CO, and a polar ice mixture with CO (CO$_P$); and finally black data shows H$_2$O, CO, CO$_P$ and \choh ices.  Ice detections were reported in \citet{Chu2020}.  }
\label{fig:BG_density}
\end{figure}

\begin{figure}[!htb]
\centering
\includegraphics[width=1\textwidth]{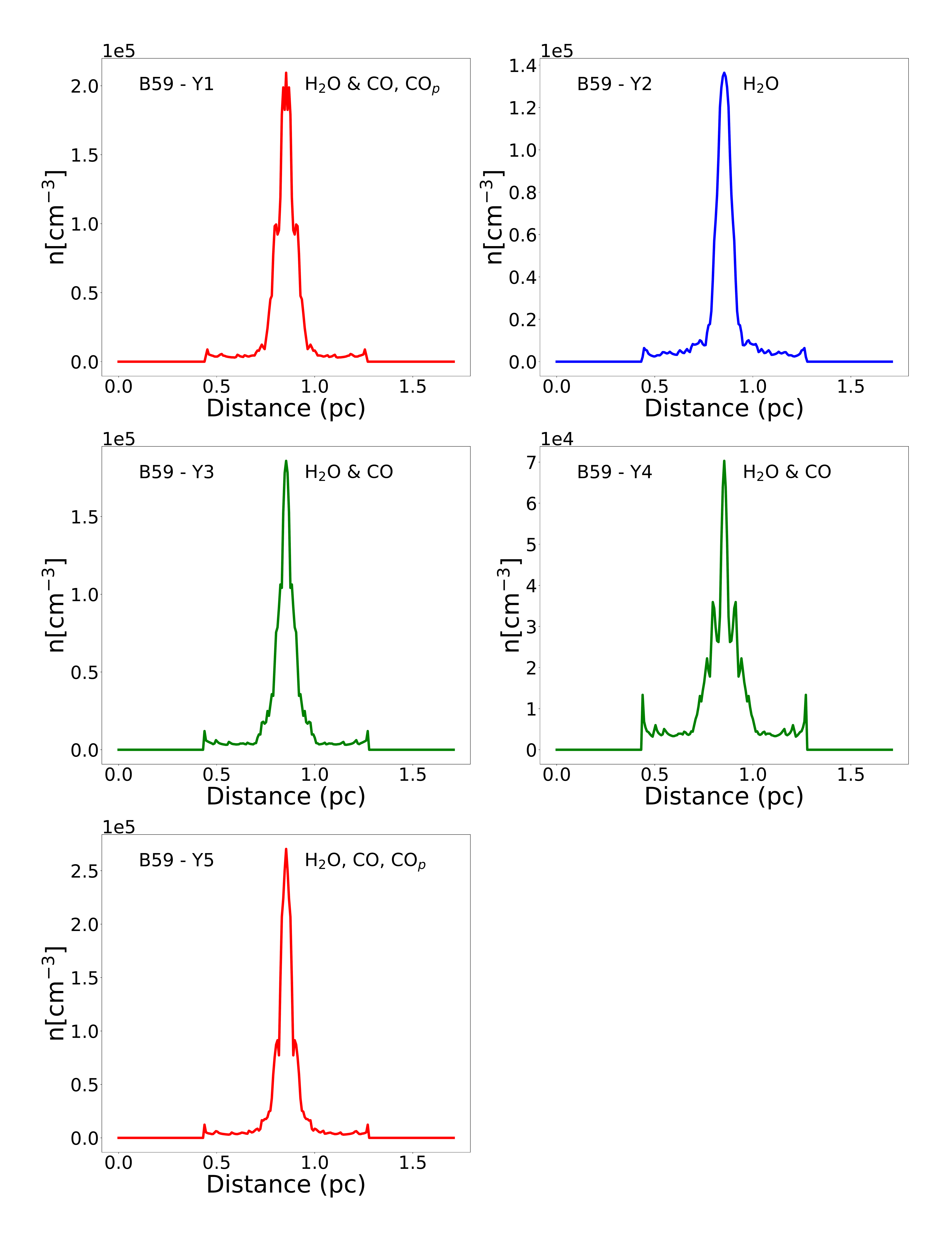}
\caption{Similar to Figure \ref{fig:BG_density} this samples lines of sight for YSOs only in B59.  Colors and axes are the same as Figure \ref{fig:BG_density}.} 
\label{fig:YSO_density}
\end{figure}

Along the lines of sight toward all thirteen ice detections we calculate discrete extinction values by taking the \Av for each voxel with the x and y coordinate corresponding to the background star or YSO and averaging the \Av with its surrounding voxels in the x and y direction, essentially as a form of binning the data further. This is to minimize artifacts from possible oversampling of the data in the 3D maps, however structures smaller than $\sim$0.01 pc are not reliable. The spatial density of total hydrogen ($\text{H} + \text{H}_2$) in n[cm$^{-3}$] is then calculated from these averaged voxel Av measurements along the lines of sight as explained in the previous Section (\ref{sec:Abel}) using the N$_{H}$/A$_V$ relationship from \citet{Predehl1995}.  Figures \ref{fig:BG_density} and \ref{fig:YSO_density} show the lines of sight for each source and are separated into background stars and YSOs.  Table \ref{tab:lines of sight} reports the maximum volume density along each line of sight with errors calculated similarly to the maximum volume density in each core. 

Table \ref{tab:Av comp} shows the total \Av along the lines of sight produced using NICER for each star and the \Av from the beam-averaged extinction map (using the beam size corresponding to each map's resolution).  The spectra for the background stars were also modeled in \citet{Chu2020} producing the extinction, A$_K$.  Using the conversion A$_V$/A$_K$=8.93 \citep{Rieke1985}, this \Av is also reported in Table \ref{tab:Av comp}.  YSO spectra could not be modeled for extinction due to a lack of suitable stellar templates for YSOs.

\begin{table*}[h!t]
\tabcolsep=0.2cm

\fontsize{9}{9}\selectfont
\centering
\caption{Ice Column Densities and Spatial Densities}
\label{tab:lines of sight}
\begin{tabular}{l|c|r|r|r|r|l}

\tableline\tableline
\multicolumn{7} {c} {Background Stars} \\

\tableline\tableline


\multicolumn{1}{l}{Source} & \multicolumn{1}{c}{Alias\footnote{Name given to identify the cloud the star samples, and whether it is a Background target (B) or YSO (Y), these match the alias in \citet{Chu2020}}} & \multicolumn{1}{c}{N(H$_2$O)\footnote{\label{chu} Ice abundances taken from \citet{Chu2020}}} & \multicolumn{1}{c}{N(CO)\footref{chu}} & \multicolumn{1}{c}{N(CO$_P$)\footref{chu}\footnote{CO$_p$ represents the long wavelength wing detected in the CO ice feature indicating a mixture with polar ice - most likely CH$_3$OH ice}}  & \multicolumn{1}{c}{N(CH$_3$OH)\footref{chu}} & \multicolumn{1}{c}{Maximum\footnote{The maximum volume density associated with a voxel with errors as the standard deviation of the volume density for surrounding voxels within the resolution of the extinction maps.  Errors are in some cases unreasonably small on their own and other errors discussed in the text will dominate (e.g. errors from the AVIATOR reconstruction, the variation in the N$_H$/A$_V$ relationship, and uncertain distances to the cores).}} \\ 

2MASS J &  & 10$^{18}$ cm$^{-2}$ & 10$^{17}$ cm$^{-2}$ &  10$^{17}$ cm$^{-2}$ & 10$^{17}$ cm$^{-2}$  &  10$^{5}$ cm$^{-3}$   \\

\tableline

17111501-2726180	&	B59-B1	&	2.45 (0.46)	&	$<$14.5	&		&		&	2.3 (0.082)	\\
18171765-0439379	&	L483-B1	&	0.17 (0.05)	&	$<$3.3	&	&	& 0.11 (0.00042) 	\\
18172690-0438406	&	L483-B2	&	5.34 (0.87)	&	12.80 (0.56)	&	5.51 (0.37)	&	3.01 (0.26)	&	3.6 (0.66)	\\
18173285-0442271	&	L483-B3	&	0.16 (0.03)	&	0.62 (0.06)	&		&	$<$1.3	&	0.55 (0.11)	\\
18174365-0438205 	&	L483-B4	&	0.19 (0.04)	&	$<$0.9	&		&	&	0.71 (0.20)	\\
19365867+0733595 	&	B335-B1	&	0.78 (0.12)	&	4.27 (0.17)	&	1.45 (0.13)	&	$<$0.7	&	1.6 (0.019)	\\
19410754+1056277 	&	L694-B2	&	2.10 (0.24)	&	11.13 (0.51)	&	4.12 (0.33)	&	2.98 (0.12)	&	1.3 (0.0054)	\\
19411163+1054416 	&	L694-B3	&	0.46 (0.07)	&	0.52 (0.26)	&	&	& 0.21 (0.00039)	\\
						
\tableline\tableline

\multicolumn{7} {c} {Young Stellar Objects} \\
\tableline\tableline
					
\multicolumn{1}{l}{Source} & \multicolumn{1}{c}{Alias} & \multicolumn{1}{c}{N(H$_2$O)} & \multicolumn{1}{c}{N(CO)} & \multicolumn{1}{c}{N(CO$_P$)}  & \multicolumn{1}{c}{N(CH$_3$OH)} & \multicolumn{1}{c}{Maximum} \\ 

2MASS J &  & 10$^{18}$ cm$^{-2}$ & 10$^{17}$ cm$^{-2}$ &  10$^{17}$ cm$^{-2}$ & 10$^{17}$ cm$^{-2}$  & 10$^{5}$ cm$^{-3}$   \\				
\tableline
17111827-2725491 	&	B59-Y1	&	2.86 (0.36)	&	8.50 (0.35)	&	2.86 (0.30)	&	$<$2.9	&	2.1 (0.16)	\\
17112153-2727417 	&	B59-Y2	&	0.52 (0.12)	&	$<$0.9	&		&	$<$1.3	&	1.4 (0.010)	\\
17112508-2724425 	&	B59-Y3	&	1.76 (0.25)	&	2.06 (0.69)	&		&		&	1.9 (0.44)	\\
17112701-2723485 	&	B59-Y4	&	1.57 (0.23)	&	1.55 (0.17)	&		&	$<$2.6	&	0.70 (0.037)	\\
17112729-2725283 	&	B59-Y5	&	1.35 (0.21)	&	5.05 (0.29)	&	2.26 (0.21)	&	$<$2.1	&	2.7 (0.11)	\\

\tableline\tableline

\end{tabular}
\end{table*}

\begin{table*}[h!t]
\tabcolsep=0.1cm
\tabletypesize{\small}
\centering
\caption{A$_V$ Comparison}
\label{tab:Av comp}
\begin{tabular}{l|c|c|c}

\tableline\tableline

Star ID & A$_V$ NICER\footnote{A$_V$ calculated using NICER with JHK and IRAC channels 1 and 2 with 1$\sigma$ errors}  & A$_V$ Beam Avg\footnote{A$_V$ calculated from the beam-averaged extinction map with the beam size corresponding to the map's resolution and with 1$\sigma$ errors reported as the standard deviation} & A$_V$ Spectrum\footnote{A$_V$ taken from the stellar spectra in \citet{Chu2020} with 1$\sigma$ errors} \\
\tableline
B59-B1 & 36.0 $\pm$ 1.8 & 31.0 $\pm$ 1.9 & 32.06 $\pm$ 0.74 \\
B59-Y1 & 49.8 $\pm$ 2.2 & 40.7 $\pm$ 1.4 & - \\
B59-Y2 & 22.6 $\pm$ 0.6 & 29.8 $\pm$ 1.2 & -  \\
B59-Y3 & 30.3 $\pm$ 0.7 & 31.2 $\pm$ 2.8 & -  \\
B59-Y4 & 14.3 $\pm$ 0.7 & 14.7 $\pm$ 1.6 & -  \\
B59-Y5 & 22.6 $\pm$ 0.6 & 45.6 $\pm$ 1.8 & -  \\
L483-B1 & 5.7 $\pm$ 1.0 & 3.2 $\pm$ 0.4 & 5.89 $\pm$ 0.39 \\
L483-B2 & 40.4 $\pm$ 2.1 & 37.0 $\pm$ 1.2 & 41.79 $\pm$ 0.83 \\
L483-B3 & 8.2 $\pm$ 1.0 & 6.3 $\pm$ 1.2 & 7.05 $\pm$ 0.27 \\
L483-B4 & 7.8 $\pm$ 1.1 & 3.3 $\pm$ 0.7 & 7.86 $\pm$ 0.33 \\
B335-B1 & 15.5 $\pm$ 1.1 & 13.3 $\pm$ 2.4 & 11.52 $\pm$ 0.21 \\
L694-B2 & 27.2 $\pm$ 1.1 & 23.1 $\pm$ 2.7 & 25.99 $\pm$ 0.24 \\
L694-B3 & 7 $\pm$ 1.1 & 7.2 $\pm$ 0.6 & 6.61 $\pm$ 0.06 \\

\tableline\tableline

\end{tabular}
\end{table*}

The two lines of sight where \choh ice is detected have a maximum volume density of 3.6$\pm$0.66~$\times$10$^5$ cm$^{-3}$ and 1.3$\pm$0.0054~$\times$10$^5$~cm$^{-3}$ (L483-B2 and L694-B2, respectively).  Because the mixture of  CO ice with polar ice (CO$_P$) is most likely due to a mixture with \choh \citep{Cuppen2011, Penteado2015} we would expect that maximum densities would be similar to those with independent  \choh ice features.  Indeed, the background star in B335 has a strong detection of CO$_P$ and the maximum density is similar to the lines of sight with \choh (1.6$\pm$0.019$\times$10$^5$ cm$^{-3}$).  In \citet{Chu2020} the lack of the direct detection of \choh ice was explained for this target as being at an extinction that is too low for \choh to be prolific since \choh has only been detected above A$_V>20$.  In Table \ref{tab:Av comp} it appears that the total \Av along the line of sight toward B335-B1 is overestimated with both the NICER measurement and beam-averaged measurement compared to the directly modeled spectrum. This could mean that the volume density is an overestimate and not dense enough to form \choh as abundantly as toward L483-B2 and L694-B2. All of the other lines of sight toward background stars have \Av measurements that agree within 3$\sigma$ of the spectral models.  

Along two other lines of sight toward YSOs where polar CO ice was detected, the maximum volume density reached is also high (2.1$\pm$0.16$\times$10$^5$ and 2.7$\pm$0.11$\times$10$^5$ cm$^{-3}$) and would also indicate that there may be an abundance of \choh ice.  In \citet{Chu2020} the upper limits of \choh were high (Table \ref{tab:lines of sight}) and thus the ice was possibly missed due to sensitivity limits.  We cannot determine if the volume densities are also overestimated because the YSO spectra could not be modeled and there are some discrepancies between the NICER \Av and the beam-averaged A$_V$ because the NICER \Av is unreliable if there remains an envelope or disk around the YSO. The low sensitivities of the ice measurements in \citet{Chu2020} for B59-B1 and B59-Y3 also mean that the high maximum densities could allow for the presence of ices beyond our detection limits.  These sensitivity constraints in B59 also explain how there is no apparent differentiation on the location where different ices have formed in the core (Figure \ref{fig:radprofile}) and potentially lines of sight that sample denser parts of the core do indeed have other undetected ices. It is noteworthy that B59 has high overall maximum densities along the different lines of sight and so another possibility is that shocks or radiation fields from the YSOs have altered the ice abundances making them lower despite dense environments. 


We can conclude that \choh appears to only form above densities of $\sim$1.0$\times$10$^5$ cm$^{-3}$. The fraction of the total core that has densities above 1.0$\times$10$^5$ \cm are 0.2\% (L63), 0.4\% (B59), 2.1\% (B335), 0.7\% (L483), and 0.3\% (L694-2) meaning \choh exists in a very small fraction of the cores. 



The relationship of the extinction for each point along the line of sight toward the ices as a function of distance from the highest density region is shown in Figure \ref{fig:radprofile}.  In L483 it is clear that the \choh ice forms much closer to the densest region at a distance of  $\sim$0.1 pc whereas CO and \hto form past $\sim$0.2 pc.  In L694-2 the \choh is also at a distance of about 0.1 pc from the peak density.  In B335 the core is smaller and the ice features are detected at a closer distance to the core's densest region.  In B59 there is little separation in the distances where different sets of ices are found again probably due to the sensitivity in ice detections.


\subsubsection{Uncertainty in Relation Between N$_H$ and A$_V$}\label{sec:Nh Av}

Several studies have calculated the N$_{H}$/A$_V$ ratio where N$_{H}$ is the total number of hydrogen atoms and molecules (H and H$_2$). There are variations due to the methods used for calculating the ratio and the regions of the sky used to develop the relation.  Using two X-ray binaries with two extended sources GCX and Cas A, \citet{Reina1973} were the first to derive the relation N$_{H}$/A$_V$=1.85$\times$10$^{21}$ cm$^{-2}$ mag$^{-1}$.  Later \citet{Gorenstein1975} used independent optical extinction and column density measurements for several supernova remnants (SNRs) to obtain  N$_{H}$/A$_V$=(2.22$\pm$0.14)$\times$10$^{21}$ cm$^{-2}$ mag$^{-1}$ which was very close to the value later found in \citet{Guver2009}.  A survey of the column densities of H I toward 100 stars in \citet{Bohlin1978} shows a smaller relation of N$_{H}$/A$_V$=9.4$\times$10$^{20}$ cm$^{-2}$ mag$^{-1}$ and is typically used to describe the diffuse interstellar medium. In \citet{Predehl1995} X-ray point sources from ROSAT observations and four SNRs produced the ratio N$_{H}$/A$_V$=(1.79$\pm$0.03)$\times$10$^{21}$ cm$^{-2}$ mag$^{-1}$ and is a very common relationship to use. \citet{Hotzel2002} later considered column densities that probed the dense cloud NGC2024 IRS2 \citep{Lacy1994} and found the relation was similar at N$_{H}$/A$_V$=1.9$\times$10$^{21}$ cm$^{-2}$ mag$^{-1}$.  The uncertainties were fairly large because it was based on a single source but compared to the diffuse ISM value in \citet{Bohlin1978} it is suggested that the ratio may increase in dense cloud cores.  Since all of these estimates only vary by a factor of $\sim$2 we decided to adopt the relation in \citet{Predehl1995}, which is similar to the median value from these studies.  The errors on this relation are also very small and the same relation was adopted for some analysis in \citet{Chu2020} making for easier comparisons in this work.

\section{Discussion} \label{sec:discussion}

Previous work by \citet{Roy2014} and \citet{Hasenberger2020} utilized the inverse abel transform to construct three dimensional maps of the dense starless cores B68 and L1689B.  They start with column density maps using data from the \textit{Herschel} and \textit{Planck} satellites.  For B68 they find a peak volume density of (3.8$\pm$0.3)$\times$10$^5$ cm$^{-3}$ and (3.7$\pm$0.3)$\times$10$^5$ cm$^{-3}$  \citep[][respectively]{Roy2014, Hasenberger2020}.  This is similar to the central density of 3.4$^{+0.9}_{-2.5}\times$10$^5$ cm$^{-3}$ found in \citet{Nielbock2012} using radiative transfer models.  For L1689B the peak density values from \citet{Roy2014, Hasenberger2020} are (9.5$\pm$1.0)$\times$10$^5$ cm$^{-3}$ and (9.5$\pm$0.5)$\times$10$^5$ cm$^{-3}$, respectively.  Both of these maximums are similar to those found for the starless collapsing cores L694-2 and L63.  The other cores in our sample have higher maximum densities of $\sim$1.2-1.7$\times$10$^6$ cm$^{-3}$ and are star forming.  B59 has the highest maximum density and also has several YSOs already formed in later Class II stages. Since B68 is not yet collapsing and has the lowest maximum density, the trend of increasing maximum density as the cores evolve into star forming cores is clear. 


The trend for the local densities where different ices can form is not as straightforward.  As mentioned in the introduction, \hto ice can form at low extinctions typically, and above an extinction of A$_V\sim$5 CO freezes out.  It would be expected that lower local densities would have \hto and at higher densities CO and \choh would be formed.  Our results show that this is generally the case but the lack of ice detections due to sensitivity limits complicates evidence of this trend.

In \citet{Chu2020} it is shown that only a small amount of CO is frozen out ($\leq$15\%) but $\sim$30\% of the CO ice is mixed with \choh ice in two cores (L483 and L694-2).  This implies that this mixture traces some of the densest parts of the core.  Our results confirm this showing that less than $\sim$2\% of the volume of L4823 and L694-2 is dense enough for \choh formation to occur.  This small region of the core is however sufficient in converting CO into \choh efficiently and allowing for complex organic molecule growth.

\section{Summary} \label{sec:summary}

Five dense molecular cores were studied to constrain the total hydrogen volume densities where ice formation occurs and understand how densities vary in different protostellar environments.  Using cores with a large population of background stars, near infrared photometry was used to develop extinction maps with very high spatial resolutions.  Implementing the new AVIATOR algorithm these maps were projected into three-dimensional space allowing for measurements of the maximum density reached in each core.  We find that the maximum density increases from cores that have not yet begun collapse to those that have infalling material and finally the highest densities are in star forming cores.  The 3D maps also provided a way to determine the density required along the lines of sight where different ices form.  We are particularly interested in the density at which \choh forms, since that may indicate the initiation of more complex organic molecular growth. It is apparent that the \choh ice forms at higher densities than other ices above 1$\times$10$^5$ cm$^{-3}$.  For the cores where \choh was detected these densities are only reached in less than $\sim$2\% of the total core hence \choh and other complex organic molecules only trace very small dense regions.

Our method demonstrates a way to use NIR photometry to determine volume densities without the dependency on submillimeter emission data or radiative transfer models that introduce different assumptions.  Extinction maps however also have drawbacks near the core center due to a lack of background stars.  With the upcoming \textit{James Webb Space Telescope} fainter targets will be observable even in the most extincted regions which should improve these measurements of the densest cores.

{\it Acknowledgements} -- We greatly appreciate the comments from the reviewer who helped significantly improve the discussion on various errors to consider.  We thank Jason Chu for helpful discussions during data and error analysis and helpful revisions from Yvonne Pendleton.  We also thank Birgit Hasenberger who was essential in learning and implementing the AVIATOR algorithm. This material is based upon work supported by the National Aeronautics and Space Administration (Grant NAS5-02105) and by the Spitzer Space Telescope (PID 11028).  When the data reported here were acquired, UKIRT was supported by NASA and operated under an agreement among the University of Hawaii, the University of Arizona, and Lockheed Martin Advanced Technology Center; operations were enabled through the cooperation of the East Asian Observatory.  We also thank the Soroptimist International Founder Region Fellowship for Women for their generous contribution supporting this work.  The authors recognize that the summit of Maunakea has always held a very significant cultural role for the indigenous Hawaiian community. We are thankful to have the opportunity to use observations from this mountain.

\bibliographystyle{aasjournal}
\bibliography{references.bib}
\end{document}